\documentclass[aps,prl,twocolumn,showpacs]{revtex4-1}

\usepackage{amsmath, amsthm, amssymb}
\usepackage{amsfonts}
\usepackage{graphicx}
\usepackage{subfigure}
\usepackage{color}
\usepackage{times}
\usepackage{float}
\usepackage{hyperref}

\hypersetup{pdfpagemode=UseNone}

\hypersetup{
        colorlinks=true,
}

\makeatletter

\pdfpageheight\paperheight
\pdfpagewidth\paperwidth

\@ifundefined{textcolor}{}{%
 \definecolor{BLACK}{gray}{0}
 \definecolor{WHITE}{gray}{1}
 \definecolor{RED}{rgb}{1,0,0}
 \definecolor{GREEN}{rgb}{0,1,0}
 \definecolor{BLUE}{rgb}{0,0,1}
 \definecolor{CYAN}{cmyk}{1,0,0,0}
 \definecolor{MAGENTA}{cmyk}{0,1,0,0}
 \definecolor{YELLOW}{cmyk}{0,0,1,0}
}

\makeatother

\begin{document}

\title{Topological Yu-Shiba-Rusinov chain in monolayer transition-metal dichalcogenide superconductors}

\author{Junhua Zhang}

\affiliation{Department of Physics \& Astronomy, University of California, Riverside,
CA 92521, USA}

\author{Vivek Aji}

\affiliation{Department of Physics \& Astronomy, University of California, Riverside,
CA 92521, USA}

\date{\today}
\begin{abstract}
Monolayers of transition-metal dichalcogenides (TMDs) are two-dimensional
materials whose low energy sector consists of two inequivalent valleys.
The valence bands have a large spin splitting due to lack of inversion symmetry and strong spin-orbit
coupling. Furthermore the spin is polarized up in one valley
and down in the other (in directions perpendicular to the two-dimensional
crystal). We focus on lightly hole-doped systems where the Fermi surface
consists of two disconnected circles with opposite spins. For both
proximity induced and intrinsic local attractive interaction induced
superconductivity, a fully gapped intervalley pairing state is favored
in this system, which is an equal superposition of the singlet and
the m=0 triplet for the lack of centrosymmetry. We show that a ferromagnetically
ordered magnetic-adatom chain placed on a monolayer TMD superconductor
provides a platform to realize one-dimensional topological superconducting
state characterized by the presence of Majorana zero modes at its
ends. We obtain the topological phase diagram and show that the topological
superconducting phase is affected not only by the adatom spacing and
the direction of the magnetic moment, but also by the orientation
of the chain relative to the crystal. 
\end{abstract}
\maketitle

\paragraph*{Introduction.--}

Monolayers of transition-metal group-VI dichalcogenides (TMDs), $\mathrm{MX_{2}}$
($\mathrm{M=Mo,\ W}$ and $\mathrm{X=S,\ Se,\ Te}$), are direct band-gap
semiconductors, which have two-dimensional (2D) hexagonal crystal
structure \cite{novoselov2005b,splendiani2010,mak2010,radisavljevic2011}.
Similar to graphene, the low-energy physics involves multiple valleys
in momentum space. There are two significant differences from graphene
and other graphene-like materials pertinent to the discussion of topological
superconductivity. First is the lack of inversion symmetry and second is the existence of a strong spin-orbit
coupling (SOC) originating from the $d$-orbitals of the heavy metal
atoms. These result in  1) a large energy gap ($\sim1.5$ eV) between the conduction
and valence bands as opposed to Dirac nodes, and 2) a Zeeman-like spin splitting ($0.15-0.5$
eV) of the valence bands, with spins polarized perpendicular to the
plane but in opposite directions between upper and lower valence bands
and between different valleys. The unique electronic structure of
monolayer TMDs, i.e., the spin degrees of freedom are locked with
the valley degrees of freedom, has triggered intensive research on
this class of 2D materials recently, e.g. a possible platform for
spintronics and valleytronics applications \cite{xiao2012,Mak27062014}.

\begin{figure}[t]
\hfill{}\includegraphics[width=3.3in]{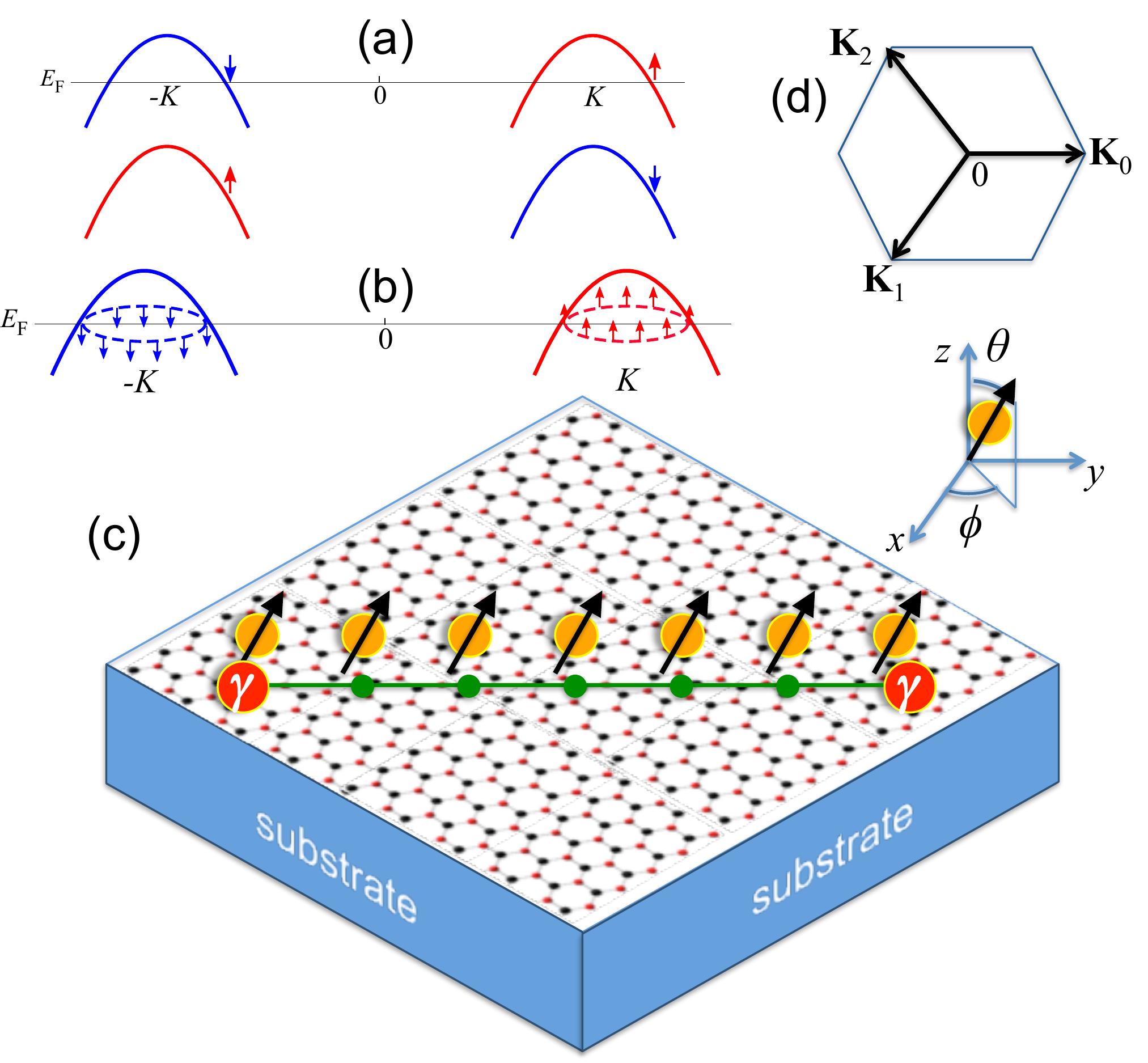}\hfill{}\caption{(Color online) Schematic illustrations of (a) the lightly-hole doped
system and (b) its disconnected Fermi surface pieces with opposite
spin directions. (c) Schematic setup of the proposed magnetic moments
forming a ferromagnetic chain on the monolayer TMD superconductor. We
show that Majorana zero modes $\gamma$ can be realized at the ends
of the YSR chain induced by the magnetic moments. (d) Schematic illustration
of the three momenta $\mathbf{K}_{n}$, $n=0,1,2$, associated with
the valley centers. \label{fig: Fig1}}
\end{figure}

The physics of spin-valley locking characterized by an Ising-type configuration
of electron spins in momentum space is maximal in the lightly hole-doped
systems. In this regime the Fermi energy crosses the upper valence
bands and is well separated from the lower valence bands as shown
in Fig. \ref{fig: Fig1}(a). The low-energy physics is characterized
by disconnected non-degenerate Fermi surfaces (FS's) in difference
valleys with opposite spin directions, as shown in Fig. \ref{fig: Fig1}(b)
. The superconducting (SC) state in this system has been studied recently
\cite{Lu2015,Saito2016,Xi2016,Zhou2015,sosenko2015arix}. The superconductivity, resulting
either from a local attractive density-density interaction or from
proximity to an $s$-wave superconductor, is characterized by inter-valley
pairing with the Cooper pair being an equal mixture of the singlet
and m=0 triplet (note that parity is no longer a good quantum number).
Since the Cooper-pair partners of opposite spin live on disconnected
FS's with spin pinned normal to the plane, dubbed Ising superconductivity,
novel properties associated with the SC state of the system in this
new regime are anticipated.

In particular, the recent efforts to search for possible platforms
to realize topological superconductivity and Majorana zero modes have
identified the need for strong SOC, time reversal breaking and superconductivity.
A relevant proposal is that of a magnetic-adatom chain placed on top
of an $s$-wave superconductor with (or effectively with) SOC \cite{Choy2011,Nadj-Perge13,klinovaja2013,braunecker2013,vazifeh2013,Pientka2013,Pientka2013b,Nakosai2013,poyhonen2014,kim2014,brydon2014,Ebisu14,LiJ2014,Peng15,Heimes15,weststrom2015,jhzhang2016,nadjperge2014,pawlak2015}.
In this scenario, the topological band arises from hybridizing the
Yu-Shiba-Rusinov (YSR) states \cite{yu1965,shiba1968,Rusinov69}
at different impurity sites. However, the previous proposals, either
based on helical spin chains \cite{Choy2011,Nadj-Perge13,braunecker2013,Pientka2013,Pientka2013b,klinovaja2013,Nakosai2013,vazifeh2013,poyhonen2014,weststrom2015}
or on superconductors with Rashba SOC \cite{brydon2014,jhzhang2016},
all require a chiral (or effectively chiral) spin texture of the electrons
coupled to the magnetic moments. While Ising spin structure lacks
chirality, TMDs do possess two essential ingredients to realize one-dimensional
(1D) topological superconductivity: (1) an odd-parity component to
the order parameter and (2) spin structure of the Cooper pair locked
perpendicular to the 2D crystal. As a result, monolayer TMD superconductor,
with ferromagnetically ordered magnetic-adatom chain placed on it,
has a large phase space for realizing Majorana zero modes making it
a possible platform for topological quantum computation 
\cite{Moore1991,Nayak1996,ReadGreen, Ivanov,TQCreview,
AliceaRev, BeenakkerReview}.

In this Rapid Communication, we study a chain of magnetic adatoms placed on the
monolayer TMD superconductors, as shown in Fig. \ref{fig: Fig1}(c),
in which the superconductivity of monolayer TMDs results either from a local
attractive density-density interaction or by proximity to an $s$-wave
superconducting substrate, e.g., $\mathrm{Nb}$, $\mathrm{NbS_{2}}$,
etc.. We show that a chain of ferromagnetically ordered magnetic impurities
on the monolayer TMD superconductors can induce a topological superconducting
phase in the YSR band. To this end, we analytically construct a tight-binding
description for the deep YSR bands, i.e., the subgap bands close to
the center of the host SC gap, to obtain an effective Hamiltonian
and calculate the topological phase diagram. We find that the topological
property is affected by the orientation of the chain on the 2D crystal
plane and the direction of the magnetic moment. When the chain is
oriented along the arm-chair directions of the hexagonal lattice which
correspond to the mirror planes in this material, the odd-parity component
vanishes due to the mirror symmetry along these directions, resulting
in vanishing topological pairing in the chain. When the magnetic moments
of the adatoms point perpendicular to the plane, the topological
pairing also vanishes since it is parallel to the characteristic SOC
direction in this material. However, for any direction of the magnetic
moments with a finite in-plane component and any orientation of the
chain other than the specific (three mirror) directions, there exists a 
large topological phase space in this system as shown in the topological
phase diagram.

\paragraph*{\textcolor{black}{Superconducting state in monolayer TMDs.}--}

The superconducting state in the lightly hole-doped monolayer TMDs,
arising either from a local attractive density-density interaction
or from proximity to an $s$-wave superconductor, is described by
the mean-field Hamiltonian ($\hbar=1$) with superconducting gap $\Delta$
\begin{equation}
\mathcal{H}_{SC}(\mathbf{k})=\sum_{\tau s}\xi_{k}d_{\tau s}^{\dagger}(\mathbf{k})d_{\tau s}(\mathbf{k})+\Delta d_{+\uparrow}^{\dagger}(\mathbf{k})d_{-\downarrow}^{\dagger}(-\mathbf{k})+h.c.\label{eq: SC_Hamiltonian}
\end{equation}
where $\xi_{k}=-k^{2}/2m-\mu$ is the valence band dispersion at chemical
potential $\mu$, $\tau=\pm$ is the valley index and $s=\uparrow,\downarrow$
is the spin index which are in the combinations as $\tau s=+\uparrow,\ -\downarrow$
due to spin-valley locking in the system. Note that we use $\mathbf{k}$
to represent momentum measured from the corresponding valley center
and $\mathbf{p}$ to represent momentum measured from the Brillouin
zone (BZ) center. We also take the notation that $d_{\tau s}^{\dagger}(\mathbf{k})$
($d_{\tau s}(\mathbf{k})$ ) creates (annihilates) a quasiparticle
with momentum $\mathbf{k}$ and spin $s$ in the valley $\tau$, and
$c_{s}^{\dagger}(\mathbf{p})$ ($c_{s}(\mathbf{p})$) creates (annihilates)
a quasiparticle with momentum $\mathbf{p}$ and spin $s$. The Hamiltonian
(\ref{eq: SC_Hamiltonian}) describes a fully gapped superconductor
with inter-valley pairing. For convenience, we take the order parameter
$\Delta$ to be real.

Transforming from the $d_{\tau s}(\mathbf{k})$ basis to the $c_{s}(\mathbf{p})$
basis, the real-space Green's function for the superconductor, defined
in the spin and particle-hole space by choosing the convention for
the Nambu spinor as $\left(\begin{array}{cccc}
c_{\uparrow}(\mathbf{r}), & c_{\downarrow}(\mathbf{r}), & c_{\downarrow}^{\dagger}(\mathbf{r}), & -c_{\uparrow}^{\dagger}(\mathbf{r})\end{array}\right)^{T}$ where $c_{s}^{\dagger}(\mathbf{r})$ ($c_{s}(\mathbf{r})$) is the
Fourier transform of $c_{s}^{\dagger}(\mathbf{p})$ ($c_{s}(\mathbf{p})$),
has the form

\begin{equation}
\hat{G}(\mathbf{r},E)=\hat{G}_{\text{even}}(\mathbf{r},E)+\hat{G}_{\text{odd}}(\mathbf{r},E),
\end{equation}
where the subscripts represent spatially even and odd parts with the
expressions 
\begin{align}
\hat{G}_{\text{even}}(\mathbf{r},E)=\frac{1}{2} & \Bigl\{\left[\mathcal{I}_{0}^{+}(\mathbf{r},E)+\mathcal{I}_{0}^{-}(\mathbf{r},E)\right]\left(E+\Delta\hat{\rho}_{x}\right)\nonumber \\
 & +\left[\mathcal{I}_{1}^{+}(\mathbf{r},E)+\mathcal{I}_{1}^{-}(\mathbf{r},E)\right]\hat{\rho}_{z}\Bigr\},\label{eq: even_G}
\end{align}
\begin{align}
\hat{G}_{\text{odd}}(\mathbf{r},E)=\frac{1}{2} & \Bigl\{\left[\mathcal{I}_{0}^{+}(\mathbf{r},E)-\mathcal{I}_{0}^{-}(\mathbf{r},E)\right]\left(E+\Delta\hat{\rho}_{x}\right)\nonumber \\
 & +\left[\mathcal{I}_{1}^{+}(\mathbf{r},E)-\mathcal{I}_{1}^{-}(\mathbf{r},E)\right]\hat{\rho}_{z}\Bigr\}\hat{\sigma}_{z}.\label{eq: odd_G}
\end{align}
The Pauli matrices $\hat{\sigma}$'s and $\hat{\rho}$'s operate in
spin and particle-hole space, respectively. $\mathcal{I}_{0}^{\pm}(\mathbf{r},E)\equiv\frac{1}{3}\sum_{n=0,1,2}e^{\pm i\mathbf{K}_{n}\cdot\mathbf{r}}I_{0}(\mathbf{r},E)$
and $\mathcal{I}_{1}^{\pm}(\mathbf{r},E)\equiv\frac{1}{3}\sum_{n=0,1,2}e^{\pm i\mathbf{K}_{n}\cdot\mathbf{r}}I_{1}(\mathbf{r},E)$
take an average over the three pairs of valleys centered at momenta
(measured from BZ center) $\pm\mathbf{K}_{n}$, $n=0,1,2$, recovering
the three-fold rotational symmetry in the continuum model. Here we
choose the direction of $\mathbf{K}_{0}$ as the $x$ direction as
shown in Fig. \ref{fig: Fig1}(d), then $\mathbf{K}_{n}\cdot\mathbf{r}=Kr\cos(n\frac{2}{3}\pi+\varphi)$,
where $\varphi$ is the angle between $\mathbf{r}$ and $\mathbf{K}_{0}$
(the $x$ axis). $I_{0}(\mathbf{r},E)$ and $I_{1}(\mathbf{r},E)$
are defined as 
\begin{eqnarray}
I_{0}(\mathbf{r},E) & = & -\frac{N_{F}}{2\pi}\int_{-D}^{D}d\xi\int_{-\pi}^{\pi}d\phi_{\mathbf{k}}\frac{e^{ik(\xi)r\cos\phi_{\mathbf{k}}}}{\xi^{2}+\Delta^{2}-E^{2}},\\
I_{1}(\mathbf{r},E) & = & -\frac{N_{F}}{2\pi}\int_{-D}^{D}d\xi\int_{-\pi}^{\pi}d\phi_{\mathbf{k}}\frac{e^{ik(\xi)r\cos\phi_{\mathbf{k}}}\xi}{\xi^{2}+\Delta^{2}-E^{2}},
\end{eqnarray}
where $D$ is an energy cut-off, $\phi_{\mathbf{k}}=\arctan(k_{y}/k_{x})$,
$k(\xi)=k_{F}+\xi/v_{F}$ with Fermi momentum $k_{F}$ and Fermi velocity
$v_{F}$, and $N_{F}$ is the density of states at the Fermi energy.
The analytical results for the above integrals are evaluated in the
limit $D\rightarrow\infty$ and are presented in the Supplemental
Material. The spatially-odd property of $\hat{G}_{\text{odd}}$ is
due to the fact that $\mathcal{I}_{0(1)}^{\pm}(-\mathbf{r})=\mathcal{I}_{0(1)}^{\mp}(\mathbf{r})$.

Equation (\ref{eq: odd_G}) shows that the odd-parity part has a spin
structure with a characteristic direction perpendicular to the plane
($z$ direction), originating from the intrinsic SOC. The existence
of an odd-parity component with a specific spin structure is the key
to the presence of topological superconductivity, as shown later.
When $\mathbf{r}$ is perpendicular to each of the $\mathbf{K}_{n}$,
$\mathbf{r}\perp\mathbf{K}_{n}$ $n=0,1,2$, the odd-parity part vanishes.
This corresponds to three special directions parallel to the arm-chair
edges of the hexagonal lattice about which mirror symmetry is preserved.

\paragraph*{Yu-Shiba-Rusinov chain.--}

We consider a chain of magnetic adatoms placed on superconducting
monolayer TMDs. The magnetic moments $\mathbf{S}_{j}=S(\sin\theta_{j}\cos\phi_{j},\ \sin\theta_{j}\sin\phi_{j},\ \cos\theta_{j})$
are located at positions $\mathbf{R}_{j}$ separated by equal spacing $a$
with classical spin $S$ pointing in the direction $(\theta_{j},\ \phi_{j})$
as shown in Fig. 1(c). Here we consider a ferromagnetic ordering of
the magnetic moments: $\theta_{j}=\theta$ and $\phi_{j}=\phi$. The
impurity magnetic moment couples to the quasiparticles in the host
superconductor through $\mathcal{H}_{imp}=\int d\mathbf{r}\psi^{\dagger}(\mathbf{r})\mathcal{H}_{imp}(\mathbf{r})\psi(\mathbf{r})$
where $\psi(\mathbf{r})=\left(\begin{array}{cccc}
c_{\uparrow}(\mathbf{r}), & c_{\downarrow}(\mathbf{r}), & c_{\downarrow}^{\dagger}(\mathbf{r}), & -c_{\uparrow}^{\dagger}(\mathbf{r})\end{array}\right)^{T}$ is the Nambu spinor at position $\mathbf{r}$, and

\begin{equation}
\mathcal{H}_{imp}(\mathbf{r})=-J\sum_{j}\mathbf{S}_{j}\cdot\hat{\boldsymbol{\sigma}}\delta(\mathbf{r}-\mathbf{R}_{j}),
\end{equation}
where $J$ is the exchange coupling ($J>0$). In order to find the band structure
of the chain of YSR states induced by the magnetic impurities, we
solve the Schrodinger equation 
\begin{equation}
\left[\mathcal{H}_{SC}+\mathcal{H}_{imp}(\mathbf{r})\right]\psi(\mathbf{r})=E\psi(\mathbf{r}).
\end{equation}
In terms of the Green's function for the superconductor $\hat{G}(\mathbf{p},E)=\left[E-\mathcal{H}_{SC}(\mathbf{p})\right]^{-1}$
and transforming to momentum space, the equation can be rewritten
as 
\begin{equation}
\psi(\mathbf{p})=-J\sum_{j}\hat{G}(\mathbf{p},E)\left(\mathbf{S}_{j}\cdot\hat{\boldsymbol{\sigma}}\right)e^{-i\mathbf{p}\cdot\mathbf{R}_{j}}\psi(\mathbf{R}_{j}).
\end{equation}
Transforming back to real space and considering the spinor at site
$i$ on the left-hand side, the chain equation becomes 
\begin{equation}
\psi(\mathbf{R}_{i})=-J\sum_{j}\hat{G}(\mathbf{R}_{i}-\mathbf{R}_{j},E)\left(\mathbf{S}_{j}\cdot\hat{\boldsymbol{\sigma}}\right)\psi(\mathbf{R}_{j}),
\end{equation}
where $\hat{G}(\mathbf{R}_{i}-\mathbf{R}_{j},E)=\int\frac{d\mathbf{p}}{(2\pi)^{2}}e^{i\mathbf{p}\cdot\left(\mathbf{R}_{i}-\mathbf{R}_{j}\right)}\hat{G}(\mathbf{p},E)$.

To solve for the bound-state spectrum, separating the on-site and
inter-site terms and using the short-hand notation $\psi_{i}\equiv\psi(\mathbf{R}_{i})$,
the chain equation takes the form: 
\begin{equation}
\hat{M}(E)\psi_{i}+\sum_{j\neq i}\hat{M}_{ij}(E)\psi_{j}=0,
\end{equation}
with the expressions for the on-site and inter-site matrices 
\begin{eqnarray}
\hat{M}(E) & = & 1+J\hat{G}(\mathbf{0},E)\left(\mathbf{S}_{i}\cdot\hat{\boldsymbol{\sigma}}\right),\label{eq: on-site}\\
\hat{M}_{ij}(E) & = & J\hat{G}(\mathbf{R}_{ij},E)\left(\mathbf{S}_{j}\cdot\hat{\boldsymbol{\sigma}}\right),\label{eq: inter-site}
\end{eqnarray}
where $\mathbf{R}_{ij}\equiv\mathbf{R}_{i}-\mathbf{R}_{j}$ for $j\neq i$,
and $\hat{G}(\mathbf{0},E)=\int\frac{d\mathbf{p}}{(2\pi)^{2}}\hat{G}(\mathbf{p},E)=-\pi N_{F}\frac{E+\Delta\hat{\rho}_{x}}{\sqrt{\Delta^{2}-E^{2}}}.$
The YSR bound state induced by a single magnetic impurity is determined
by $\hat{M}(E)\psi_{i}=0$, giving rise to the eigen energies $E_{\pm}=\pm\Delta\frac{1-\tilde{J}^{2}}{1+\tilde{J}^{2}}$
in terms of the dimensionless exchange coupling $\tilde{J}\equiv J\pi N_{F}S$,
and the eigen spinors 
\begin{equation}
\psi_{+}\sim\left(\begin{array}{c}
\chi_{\uparrow}\\
\chi_{\uparrow}
\end{array}\right),\ \ \psi_{-}\sim\left(\begin{array}{c}
\chi_{\downarrow}\\
-\chi_{\downarrow}
\end{array}\right),
\end{equation}
\begin{equation}
\chi_{\uparrow}=\left(\begin{array}{c}
\cos\frac{\theta}{2}\\
e^{i\phi}\sin\frac{\theta}{2}
\end{array}\right),\ \ \chi_{\downarrow}=\left(\begin{array}{c}
e^{-i\phi}\sin\frac{\theta}{2}\\
-\cos\frac{\theta}{2}
\end{array}\right),
\end{equation}
up to a normalization constant. These eigen spinors $\psi_{\pm}$
are used as local basis when deriving the effective Hamiltonian for
the YSR chain.

\paragraph*{Tight-binding description.--}

To analytically construct a tight-binding description for the YSR
chain \cite{Pientka2013,brydon2014,jhzhang2016}, we take the deep
and shallow band approximations: Consider the YSR band is close to
the center of the host gap, i.e., $\tilde{J}\sim1$, and its bandwidth
is small compared to the host gap, i.e., $E\ll\Delta$, which requires
a large spacing $a$ between the adatoms. In these approximations,
we linearize $\hat{M}(E)$ with respect to $E$: $\hat{M}(E)\approx\hat{M}_{0}-\hat{M}_{1}\cdot E$
, and set $E\rightarrow0$ in the inter-site matrix $\hat{M}_{ij}(E\rightarrow0)$.
Then the tight-binding description is given by

\begin{equation}
\sum_{j}\hat{H}_{ij}\psi_{j}=E\psi_{i}
\end{equation}
with the on-site and inter-site operators being $\hat{H}_{ii}=\hat{M}_{1}^{-1}\hat{M}_{0}$
and $\hat{H}_{ij}=\hat{M}_{1}^{-1}\hat{M}_{ij}(E\rightarrow0)$, $j\neq i$,
respectively. The explicit expressions for $\hat{M}_{0}$, $\hat{M}_{1}$,
and $\hat{M}_{ij}(E\rightarrow0)$ are given in the Supplemental Material.

\paragraph*{Effective BdG Hamiltonian.--}

By projecting $\hat{H}_{ij}$ onto the local YSR basis $(\psi_{+},\ \psi_{-})^{T}$,
we obtain the effective Bogoliubov-de Gennes (BdG) Hamiltonian as
\begin{equation}
H_{\text{eff}}(i,j)=\left[\begin{array}{cc}
h_{\text{eff}}(\mathbf{R}_{ij})+b_{\text{eff}}(\mathbf{R}_{ij}) & \Delta_{\text{eff}}(\mathbf{R}_{ij})\\
\Delta_{\text{eff}}^{*}(\mathbf{R}_{ji}) & -h_{\text{eff}}(\mathbf{R}_{ij})+b_{\text{eff}}(\mathbf{R}_{ij})
\end{array}\right],
\end{equation}
where 
\begin{equation}
h_{\text{eff}}(\mathbf{R}_{ij})=\epsilon_{0}\delta_{ij}+\frac{1}{2\tilde{J}}JS\Delta^{2}\left[\mathcal{I}_{0}^{+}(\mathbf{R}_{ij})+\mathcal{I}_{0}^{-}(\mathbf{R}_{ij})\right],\label{eq: hopping}
\end{equation}
\begin{equation}
b_{\text{eff}}(\mathbf{R}_{ij})=\frac{1}{2\tilde{J}}JS\Delta^{2}\left[\mathcal{I}_{0}^{+}(\mathbf{R}_{ij})-\mathcal{I}_{0}^{-}(\mathbf{R}_{ij})\right]\cos\theta,\label{eq: Zeeman-like}
\end{equation}
\begin{equation}
\Delta_{\text{eff}}(\mathbf{R}_{ij})=\frac{-1}{2\tilde{J}}JS\Delta\left[\mathcal{I}_{1}^{+}(\mathbf{R}_{ij})-\mathcal{I}_{1}^{-}(\mathbf{R}_{ij})\right]e^{-i\phi}\sin\theta,\label{eq: pairing}
\end{equation}
with the on-site energy $\epsilon_{0}=\Delta(1-\tilde{J})/\tilde{J}$,
i.e., the effective chemical potential of the YSR band. For $\tilde{J}\sim1$,
$\epsilon_{0}\sim0$, the YSR band is close to the center of the host
gap. Here the integrals $\mathcal{I}_{0(1)}^{\pm}$ are evaluated
for $E\rightarrow0$ and considered vanishing for $i=j$.

The effective pairing term has the spatial property: $\Delta_{\text{eff}}(-\mathbf{R}_{ij})=-\Delta_{\text{eff}}(\mathbf{R}_{ij})$,
since it is arising from the odd-parity component $\hat{G}_{\text{odd}}$
of the bulk superconductor. This describes a $p$-wave pairing of
the spinless fermions in 1D, reminiscent of the Kitaev chain \cite{Kitaev2001},
despite the long-range nature of the inter-site hopping and pairing.
Clearly, when the magnetic moments point perpendicular to the plane
(in $z$ direction), $\theta=0,\pi$, the effective pairing vanishes
indicating that the subgap band is not superconducting in this case.
However, as long as the magnetic moment has a finite in-plane component,
the subgap band can become superconducting. Besides the even-parity
hopping $h_{\text{eff}}(\mathbf{R}_{ij})$, we also observe that a
polarization component along the $z$ axis induces an odd-parity inter-site
term $b_{\text{eff}}(\mathbf{R}_{ij})$. Its presence is also due
to the odd-parity component $\hat{G}_{\text{odd}}$ in the bulk superconductor.
$b_{\text{eff}}(\mathbf{R}_{ij})$ is analogous to the response to
an applied magnetic field.

\paragraph*{Topological superconducting phase.--}

To study the topological property, transforming the BdG Hamiltonian
to momentum space, we obtain 
\begin{equation}
H_{\text{eff}}(p)=\left[\begin{array}{cc}
h_{\text{eff}}(p)+b_{\text{eff}}(p) & \Delta_{\text{eff}}(p)\\
\Delta_{\text{eff}}^{*}(p) & -h_{\text{eff}}(p)+b_{\text{eff}}(p)
\end{array}\right].\label{eq: BdG_momentum_space}
\end{equation}
The details are given in the Supplemental Material. The Hamiltonian is
in the symmetry class D \cite{altland1997,schnyder2008,kitaev2008},
and thus is characterized by the $Z_{2}$ topological invariant $\mathcal{M}$
\cite{Kitaev2001}:

\begin{equation}
\mathcal{M}=\mathrm{sgn}\left[h_{\text{eff}}(0)h_{\text{eff}}(\pi/a)\right].
\end{equation}
The system is in the topological superconducting phase when $\mathcal{M}=-1$,
whereas $\mathcal{M}=+1$ indicates a non topological phase. We obtain
the topological phase diagram by calculating $\mathcal{M}$.

\begin{figure}[t]
\hfill{}\includegraphics[width=3.3in]{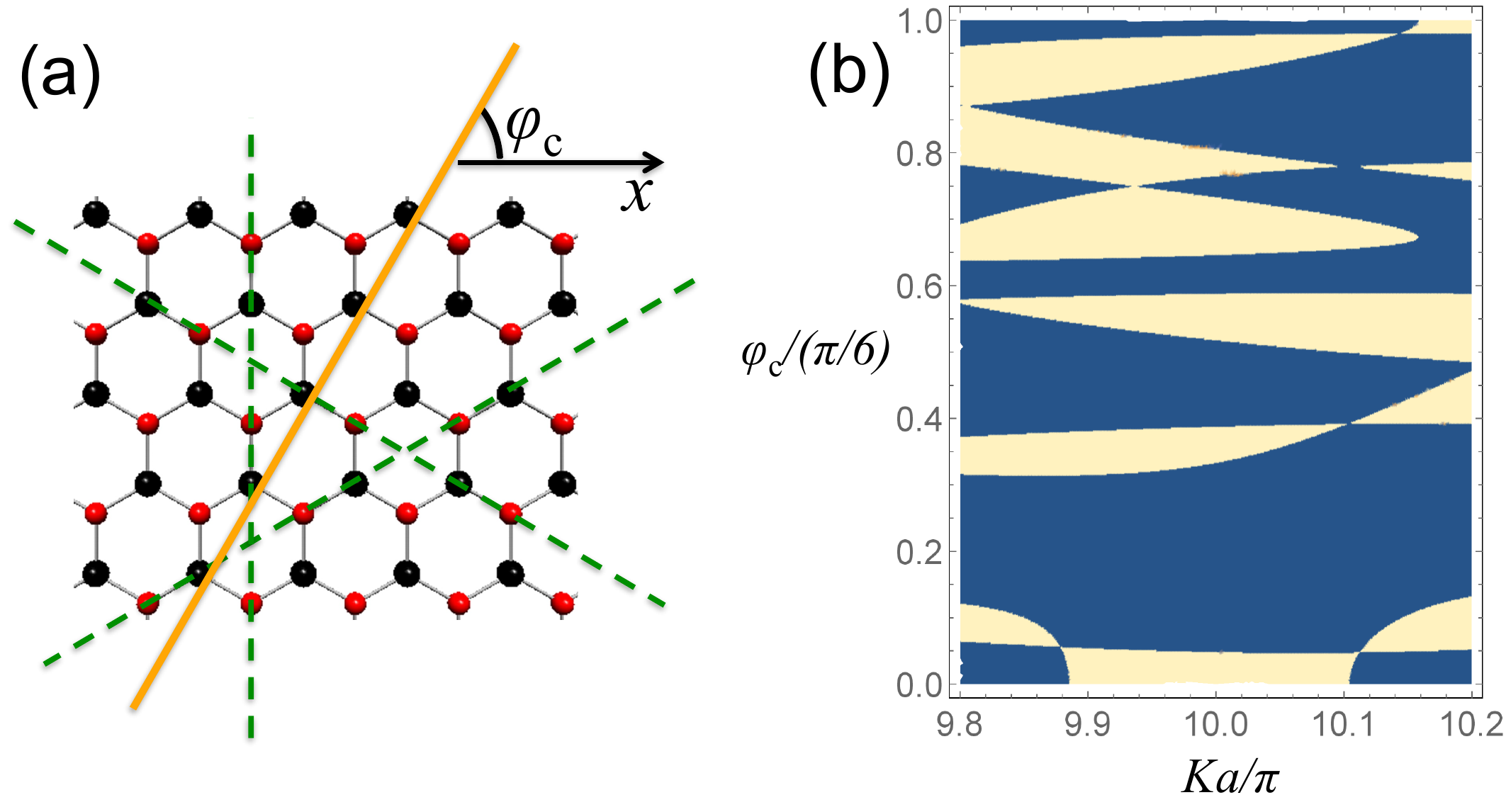}\hfill{}\caption{(Color online) (a) Schematic illustration of the chain orientation
relative to the 2D crystal structure characterized by the angle $\varphi_{c}$.
The dashed lines correspond to the mirror-plane directions along which
the YSR bands are gapless. (b) Calculated topological phase diagram
for the effective Hamiltonian (\ref{eq: BdG_momentum_space}) as a
function of $\varphi_{c}\in(0,\ \pi/6)$ and $Ka$ ($K=|\mathbf{K}_{n}|$)
for the magnetic moments aligning in the crystal plane, i.e., $\theta=\pi/2$.
Here we take $\tilde{J}=1$ such that $\epsilon_{0}=0$. The dark
color ($\mathcal{M}=-1$) represents the topological phase characterized
by an odd number of Majoranas at its ends, whereas the light color
($\mathcal{M}=+1$) refers to the non-topological phase. \label{fig: Fig2}}
\end{figure}

The topological phase is affected by the direction of the magnetic
moments as indicated in Eq. (\ref{eq: pairing}). A finite topological
gap requires the magnetic moments deviate from $z$ direction and
reaches its maximum when the moments lie in the plane. This is achieved
experimentally by aligning the moments using magnetic field before
cooling the system. The topological property is also affected by the
orientation of the chain, characterized by the angle $\varphi_{c}$
between the chain and the $x$ direction in the crystal plane, as
shown in Fig. \ref{fig: Fig2}(a). As discussed before, the inter-site
coupling (\ref{eq: inter-site}) through $\hat{G}(\mathbf{r},E)$
involves phase factors: $\mathbf{K}_{n}\cdot\mathbf{r}=Kr\cos(n\frac{2}{3}\pi+\varphi_{c})$
for $\mathbf{r}$ on the chain. When $\mathbf{r}\perp\mathbf{K}_{n}$
($n=0,1,2$), corresponding to the three arm-chair directions in the
hexagonal lattice as indicated by the dashed lines in Fig. \ref{fig: Fig2}(a),
$\Delta_{\text{eff}}(\mathbf{R}_{ij})$ vanishes due to the vanishing
odd-parity component $\hat{G}_{\text{odd}}$ in the bulk superconductor.
These special directions are related to the mirror planes in the system.
A finite topological gap requires the chain orientation to avoid these
special directions. The phase diagram calculated using the $Z_{2}$
topological invariant $\mathcal{M}$ is valid between two mirror planes,
e.g., $\varphi_{c}\in(-\pi/6,\ \pi/6)$ and is symmetric with respect 
to each mirror plane, as well as having threefold rotational symmetry.
Furthermore, it is symmetric for $\varphi_{c}$ and $-\varphi_{c}$.
Therefore, the calculation for the regime $\varphi_{c}\in(0,\ \pi/6)$
is sufficient as other sectors are related through symmetry. The adatom
spacing $a$ is another factor affecting the topological property
because the band structure of the YSR chain is sensitive to it.

Figure \ref{fig: Fig2}(b) shows the calculated topological phase
diagram for the effective Hamiltonian (\ref{eq: BdG_momentum_space})
as a function of $\varphi_{c}$ and $Ka$ ($K=|\mathbf{K}_{n}|$)
for the magnetic moments aligning in the crystal plane $\theta=\pi/2$. Here we take
$\tilde{J}=1$ such that the effective chemical potential is right
in the middle of the bulk gap. The dark color represents the topological
phase characterized by an odd number of Majoranas at its ends, whereas
the light color refers to the non-topological phase. We can see that there
is a large parameter space in which the chain is expected to be in
a topological phase.

\paragraph*{Conclusions.--}

In conclusion we have investigated the topological nature of a ferromagnetic
adatom chain proximally coupled to an inter-valley paired TMD superconductor.
Strong SOC and the lack of inversion symmetry in monolayer TMDs can
support a topological superconducting state in the YSR chain, as long
as the chain is not oriented along mirror planes of the 2D crystal
and the magnetic moments have a finite in-plane component. Thus, for
a large parameter regime, Majorana bound states at the end of the
chain can be realized, providing a new platform to explore topological
phases of matter.

\paragraph*{Acknowledgements.--}

This work is supported by ARO W911NF1510079.

\paragraph*{Notes added.--}

Recently, we became aware of a preprint
\cite{sharma2016arxiv} with similar conclusions.

%



\newpage

\onecolumngrid
\begin{center}

{\bf\Large{Supplemental Material for ``Topological Yu-Shiba-Rusinov chain in monolayer transition-metal dichalcogenide superconductors''}}

\end{center}
\vspace{1cm}
\setcounter{equation}{0}
\renewcommand{\theequation}{S\arabic{equation}}
\setcounter{figure}{0}
\renewcommand{\thefigure}{S\arabic{figure}}

\section*{analytical expressions of integrals $I_{0}(\mathbf{r},E)$ and $I_{1}(\mathbf{r},E)$}

We provide the analytical expressions of the integral functions defined
in Eqs. (5) and (6):

\begin{equation}
I_{0}(\mathbf{r},E)=-\frac{\pi N_{F}}{\sqrt{\Delta^{2}-E^{2}}}\mathrm{Re}\left\{ J_{0}\left[\left(k_{F}+i\xi_{0}^{-1}\right)r\right]+iH_{0}\left[\left(k_{F}+i\xi_{0}^{-1}\right)r\right]\right\} ,
\end{equation}
\begin{equation}
I_{1}(\mathbf{r},E)=-\pi N_{F}\mathrm{Im}\left\{ J_{0}\left[\left(k_{F}+i\xi_{0}^{-1}\right)r\right]+iH_{0}\left[\left(k_{F}+i\xi_{0}^{-1}\right)r\right]\right\} ,
\end{equation}
where $r=|\mathbf{r}|$, $J_{n}(z)$ and $H_{n}(z)$ are Bessel and
Struve functions of order $n$, respectively, and $\xi_{0}^{-1}\equiv\frac{\sqrt{\Delta^{2}-E^{2}}}{-v_{F}}$,
$v_{F}<0$ for the valence bands. $\xi_{0}^{-1}\approx\frac{\Delta}{-v_{F}}$
corresponds to the inverse of superconducting coherence length in
the host superconductor. Note that we take $\mathbf{r}\neq0$ here.

\section*{asymptotic forms for $I_{0}(\mathbf{r},E)$ and $I_{1}(\mathbf{r},E)$}

In order to perform Fourier transform of the Hamiltonian to momentum
space, we use the asymptotic forms of the Bessel and Struve functions
valid for large values of the argument close to the positive real
axis. This is valid in our calculation because we consider shallow
YSR band cases, i.e., the spacing between the adatoms is large $k_{F}a\gg1$.
In the limit $k_{F}r\gg1$, we can find the approximate expressions
for $I_{0}$ and $I_{1}$, to the leading order, as 
\begin{equation}
I_{0}(\mathbf{r},E)=-\frac{\pi N_{F}}{\sqrt{\Delta^{2}-E^{2}}}\sqrt{\frac{2}{\pi k_{F}r}}e^{-\xi_{0}^{-1}r}\cos\left[k_{F}r-\frac{1}{4}\pi\right],
\end{equation}
\begin{equation}
I_{1}(\mathbf{r},E)=-\pi N_{F}\sqrt{\frac{2}{\pi k_{F}r}}e^{-\xi_{0}^{-1}r}\sin\left[k_{F}r-\frac{1}{4}\pi\right].
\end{equation}

\section*{Fourier transforms of $\mathcal{I}_{0}^{\pm}(\mathbf{r},E)$ and
$\mathcal{I}_{1}^{\pm}(\mathbf{r},E)$ }

The Fourier transforms of the effective Hamiltonian can be carried
out analytically when utilizing the asymptotic expressions. For
\begin{equation}
\mathcal{I}_{0}^{\pm}(\mathbf{r},E)=\frac{1}{3}\sum_{n}e^{\pm i\mathbf{K}_{n}\cdot\mathbf{r}}I_{0}(\mathbf{r},E),\ \ \mathcal{I}_{1}^{\pm}(\mathbf{r},E)=\frac{1}{3}\sum_{n}e^{\pm i\mathbf{K}_{n}\cdot\mathbf{r}}I_{1}(\mathbf{r},E),
\end{equation}
define the Fourier transform as
\[
f(p)=\sum_{j=-\infty}^{\infty}f(aj)e^{ipaj}.
\]
Then the Fourier transforms of $\mathcal{I}_{0}^{\pm}(\mathbf{r})$
and $\mathcal{I}_{1}^{\pm}(\mathbf{r})$ to the momentum along the
chain take the form, 
\begin{align}
\mathcal{I}_{0}^{\pm}(p,E)= & -\frac{\pi N_{F}}{\sqrt{\Delta^{2}-E^{2}}}\sqrt{\frac{1}{2\pi k_{F}a}}\frac{1}{3}\sum_{n=0,1,2}\nonumber \\
 & \Biggl\{ e^{-i\frac{\pi}{4}}\left[\mathrm{Li}_{\frac{1}{2}}\left(e^{\pm iKa\cos(n\frac{2}{3}\pi+\varphi_{c})+ik_{F}a+ipa-\xi_{0}^{-1}a}\right)+\mathrm{Li}_{\frac{1}{2}}\left(e^{\mp iKa\cos(n\frac{2}{3}\pi+\varphi_{c})+ik_{F}a-ipa-\xi_{0}^{-1}a}\right)\right]\nonumber \\
 & +e^{i\frac{\pi}{4}}\left[\mathrm{Li}_{\frac{1}{2}}\left(e^{\pm iKa\cos(n\frac{2}{3}\pi+\varphi_{c})-ik_{F}a+ipa-\xi_{0}^{-1}a}\right)+\mathrm{Li}_{\frac{1}{2}}\left(e^{\mp iKa\cos(n\frac{2}{3}\pi+\varphi_{c})-ik_{F}a-ipa-\xi_{0}^{-1}a}\right)\right]\Biggl\},
\end{align}
\begin{align}
\mathcal{I}_{1}^{\pm}(p,E)= & i\pi N_{F}\sqrt{\frac{1}{2\pi k_{F}a}}\frac{1}{3}\sum_{n=0,1,2}\nonumber \\
 & \Biggl\{ e^{-i\frac{\pi}{4}}\left[\mathrm{Li}_{\frac{1}{2}}\left(e^{\pm iKa\cos(n\frac{2}{3}\pi+\varphi_{c})+ik_{F}a+ipa-\xi_{0}^{-1}a}\right)+\mathrm{Li}_{\frac{1}{2}}\left(e^{\mp iKa\cos(n\frac{2}{3}\pi+\varphi_{c})+ik_{F}a-ipa-\xi_{0}^{-1}a}\right)\right]\nonumber \\
 & -e^{i\frac{\pi}{4}}\left[\mathrm{Li}_{\frac{1}{2}}\left(e^{\pm iKa\cos(n\frac{2}{3}\pi+\varphi_{c})-ik_{F}a+ipa-\xi_{0}^{-1}a}\right)+\mathrm{Li}_{\frac{1}{2}}\left(e^{\mp iKa\cos(n\frac{2}{3}\pi+\varphi_{c})-ik_{F}a-ipa-\xi_{0}^{-1}a}\right)\right]\Biggl\},
\end{align}
where $\mathrm{Li}_{s}(z)$ is the polylogarithm function:
\[
\mathrm{Li}_{s}(z)=\sum_{n=1}^{\infty}\frac{z^{n}}{n^{s}}.
\]

\section*{Expressions for $\hat{M}_{0}$, $\hat{M}_{1}$, and $\hat{M}_{ij}(E\rightarrow0)$}

We provide the analytical expressions for $\hat{M}_{0}$, $\hat{M}_{1}$,
and $\hat{M}_{ij}(E\rightarrow0)$ in $\hat{H}_{ij}$ of Eq. (16):

\begin{equation}
\hat{M}_{0}=1-\tilde{J}\hat{\rho}_{x}\hat{\mathbf{S}}_{i}\cdot\hat{\boldsymbol{\sigma}},\ \ \ \hat{M}_{1}=\frac{\tilde{J}}{\Delta}\left(\hat{\mathbf{S}}_{i}\cdot\hat{\boldsymbol{\sigma}}\right),
\end{equation}
\begin{align}
\hat{M}_{ij}(E\rightarrow0)= & \frac{1}{2}JS\left\{ \left[\mathcal{I}_{1}^{+}(\mathbf{R}_{ij},0)+\mathcal{I}_{1}^{-}(\mathbf{R}_{ij},0)\right]\hat{\rho}_{z}+\Delta\left[\mathcal{I}_{0}^{+}(\mathbf{R}_{ij},0)+\mathcal{I}_{0}^{-}(\mathbf{R}_{ij},0)\right]\hat{\rho}_{x}\right\} \left(\hat{\mathbf{S}}_{j}\cdot\hat{\boldsymbol{\sigma}}\right)\nonumber \\
+ & \frac{1}{2}JS\left\{ \left[\mathcal{I}_{1}^{+}(\mathbf{R}_{ij},0)-\mathcal{I}_{1}^{-}(\mathbf{R}_{ij},0)\right]\hat{\rho}_{z}+\Delta\left[\mathcal{I}_{0}^{+}(\mathbf{R}_{ij},0)-\mathcal{I}_{0}^{-}(\mathbf{R}_{ij},0)\right]\hat{\rho}_{x}\right\} \hat{\sigma}_{z}\left(\hat{\mathbf{S}}_{j}\cdot\hat{\boldsymbol{\sigma}}\right),
\end{align}
where $\hat{\mathbf{S}}\equiv\mathbf{S}/S$.

\end{document}